\documentclass[aps,prl,amsmath,amssymb,reprint,superscriptaddress,twocolumn,showpacs,nobibnotes,
]{revtex4-2}

\usepackage{graphicx}
\usepackage{dcolumn}
\usepackage{amsmath}
\usepackage{amsfonts}
\usepackage{amssymb}
\usepackage{pstricks}
\usepackage{psfrag}
\usepackage{epsfig}
\usepackage{changes}
\usepackage[ansinew]{inputenc}
\usepackage[normalem]{ulem}
\usepackage{textcomp}
\usepackage{ulem}
\usepackage{bm}
\usepackage[colorlinks=true,linkcolor=blue,citecolor=blue,urlcolor=blue]{hyperref} 

\begin{document}

\title{Hydrodynamic Switching Fronts Polarize Deformable Particle Trains}

\author{Linzheng Huang}
\affiliation{State Key Laboratory for Turbulence and Complex Systems, School of Mechanics and Engineering Science, Peking University, Beijing 100871, China}

\author{Hengdi Zhang}
\affiliation{PaXini Technology (Shenzhen) Co., Ltd., Shenzhen 518100, China}

\author{Zaicheng Zhang}
\affiliation{School of Physics, Beihang University, Beijing 100191, China}

\author{Zaiyi Shen}
\email[Contact author: ]{zaiyi.shen@pku.edu.cn}

\affiliation{State Key Laboratory for Turbulence and Complex Systems, School of Mechanics and Engineering Science, Peking University, Beijing 100871, China}

\date{\today}

\begin{abstract}
We show that propagating switching fronts mediate directional state transmission and polarity selection in a passive many-body suspension. In confined trains of slipper-shaped deformable particles in Poiseuille flow, this behavior originates from directionally biased switching between neighboring particles: owing to the fore-aft asymmetry of the slipper, an upstream particle drives switching of its downstream neighbor more effectively than in the reverse direction. A local transition from an opposite-sign pair to a same-sign pair therefore launches a streamwise front that relays the inclination sign from particle to particle. A minimal coarse-grained model with local bistability and directional coupling captures front propagation and arrest. In periodic trains, the fronts coarsen into a uniformly polarized state, whereas in long open trains they arrest and leave persistent polarized domains. Our results point to local bistability and directional coupling as a route to collective polarization in passive many-body systems.
\end{abstract}

\maketitle


How local interactions generate directional state transmission, propagating fronts, and collective order is a recurring theme in many-body dynamics \cite{marchetti2013, van2003}. In ciliary arrays, local coupling organizes neighboring beat cycles into metachronal waves \cite{Vilfan2006, Meng2021, brumley2012, mesdjian2022}, and near-field hydrodynamic effects can even produce effective non-reciprocity and rapid propagation of order \cite{Hickey2023}. More generally, asymmetric couplings can drive drifting fronts and traveling ordered states \cite{You2020, Fruchart2021, Avni2025}. In these systems, directional state transmission is typically associated with activity or with couplings made asymmetric \emph{a priori}. Whether a passive many-body suspension can spontaneously support directional state transmission remains unknown.

A confined deformable particle in viscous Poiseuille flow can adopt an asymmetric off-centered slipper state \cite{Kaoui2009, Kaoui2011, dahl2015}. This state has two mirror-related branches with opposite inclination signs, reflecting a local \(\pm\) degeneracy of the particle orientation \cite{Kaoui2009, Agarwal2022}. For an isolated slipper, the selected sign is set by the initial condition and remains a single-particle property. In a train, however, neighboring particles are coupled by the disturbance flows they generate, so the local branch state is no longer determined by the background flow alone. Train formation is generic in confined suspensions of soft particles, including red blood cells, droplets, bubbles, vesicles, and capsules \cite{shen2018, claveria2016, Iss2019, Janssen2012, Beatus2006, Beatus2007, Raven2009, McWhirter2009}. Existing studies have focused mainly on train formation itself, and on how spacing and stability are selected \cite{Ghigliotti2012,champagne2010, Bryngelson2016, Bryngelson2018,lu2021, Millett2024}. Much less is known about how the inclination state carried by each particle is transmitted and reorganized once the train is established.

\begin{figure*}
\centering
\includegraphics[width=2\columnwidth]{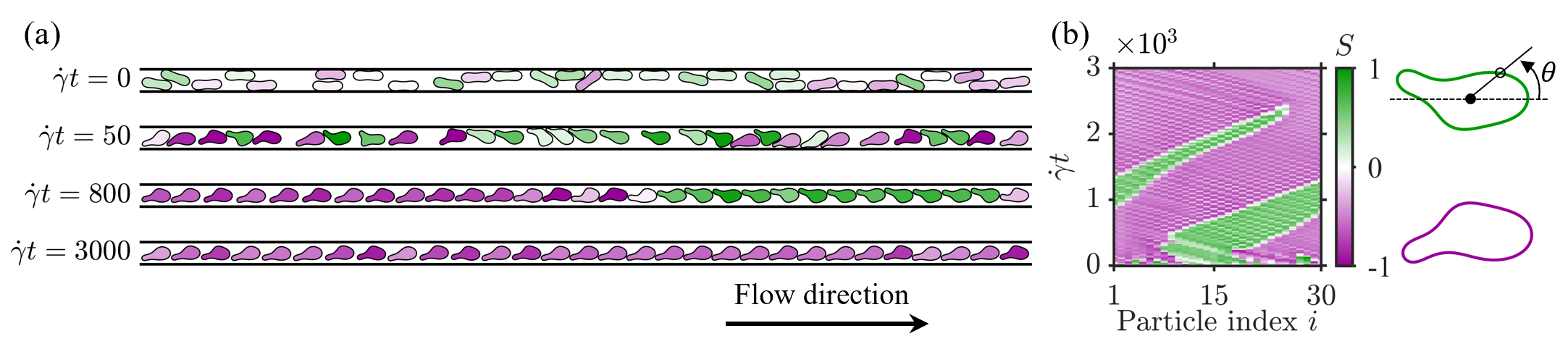}
\caption{\label{fig:fig1}
Polarity of self-organized particle trains.
(a) Time series showing the evolution of \(N=30\) deformable particles in Poiseuille flow from a random initial distribution to a single-file train and, subsequently, to a uniformly polarized inclination state. Parameters: \(Ca=10\) and \(\phi\simeq 37\%\).
(b) Corresponding space-time plot of the normalized slipper inclination \(S\), showing the propagation of switching fronts during polarization. Particle indices are assigned after the single-file train forms and are ordered along the flow direction. Colors indicate \(S\), with representative slipper shapes for \(S=\pm1\) shown on the right.}
\end{figure*}

Here we show that confined trains of deformable slippers support propagating switching fronts that mediate directional state transmission and polarity selection. The key ingredient is directionally biased switching between neighboring slippers: in confined Poiseuille flow, the influence transmitted from an upstream slipper to its downstream neighbor is more effective for switching than the reverse influence. This bias originates from the fore-aft asymmetry of the slipper and the disturbance flow it generates. As a result, a local branch switch launches a streamwise switching front that relays the inclination sign from particle to particle and drives the train toward a uniformly ordered state---a hydrodynamic ``domino'' that resolves the $\pm$ inclination degeneracy and realizes collective polarization in passive many-body systems through local bistability and directional coupling.

We study deformable particles confined between two parallel walls in Poiseuille flow [Fig.~\ref{fig:fig1}(a)] using immersed boundary-lattice Boltzmann simulations \cite{Shen2017,kruger2011}. Each particle is modeled as a two-dimensional closed spring-network membrane with a prescribed biconcave reference shape \cite{Shen2017, tsubota2006}, characterized by the reduced area \(\nu=4\pi A/P^2\), where \(A\) and \(P\) are the enclosed area and perimeter. Throughout this work, we fix \(\nu=0.65\) and the confinement ratio \(Cn=2R/W=0.8\), with \(W\) the wall spacing and \(R=\sqrt{A/\pi}\) the effective particle radius. This choice favors the formation of a single-file train of slippers. Similar behavior is also observed for nearby reduced areas and confinements, e.g., \(\nu=0.6\) and \(0.7\), and \(Cn=0.7\) and \(0.9\).

The flow is driven by a uniform body force, with periodic boundary conditions in the streamwise direction and no-slip at the walls. In the absence of particles, the velocity profile is parabolic, \(u(y)=u_m\!\left[1-4(y/W)^2\right]\), 
where \(u_m\) is the centerline velocity and \(y\) is measured from the channel centerline (walls at \(y=\pm W/2\)). We define the characteristic shear rate as \(\dot\gamma=4u_m/W\). Particle deformability is quantified by the capillary number \(Ca=\eta R^3 \dot\gamma/\kappa\), 
where \(\eta\) is the viscosity of the suspending fluid and \(\kappa\) is the membrane bending modulus. In all simulations, the Reynolds number is small (\(\sim 10^{-2}\)), so inertial effects are negligible.

\textit{Polarity of self-organized particle trains}---For a single particle in Poiseuille flow, the steady off-centered slipper state has two mirror-related branches with opposite inclination signs, selected by the initial lateral position. To quantify the inclination, we expand the particle contour in polar form,
\begin{equation}
r(\theta)=R+\sum_{n\ge1}\Big[a_n\cos(n\theta)+b_n\sin(n\theta)\Big],
\end{equation}
and define the order parameter \(s=-b_1/R\). Normalizing by its steady-state magnitude \(s_e\), we set \(S=s/s_e\), so that \(S=\pm1\) labels the two stable slipper inclinations. Here \(S>0\) denotes an upward slipper and \(S<0\) its mirror downward state [Fig.~\ref{fig:fig1}(b)].

With multiple particles present, however, the inclination state is no longer selected independently by each particle. To illustrate this collective dynamics, we initialize \(N=30\) particles at random positions [Fig.~\ref{fig:fig1}(a)], corresponding to an area fraction \(\phi=NA/(L_XL_Y)\simeq 37\%\), where \(L_X\) and \(L_Y\) are the domain sizes. At early times, the particles deform into slippers, migrate toward the centerline, and assemble into a single-file train. Once the train is established, the subsequent evolution is controlled by interparticle coupling: neighboring particles on opposite branches switch sequentially, so that the inclination sign is transmitted along the train. This front-like transmission progressively eliminates opposite-sign pairs and drives the train toward a uniformly polarized state [Fig.~\ref{fig:fig1} and Movie S1].

\begin{figure}
\centering
\includegraphics[width=1\columnwidth]{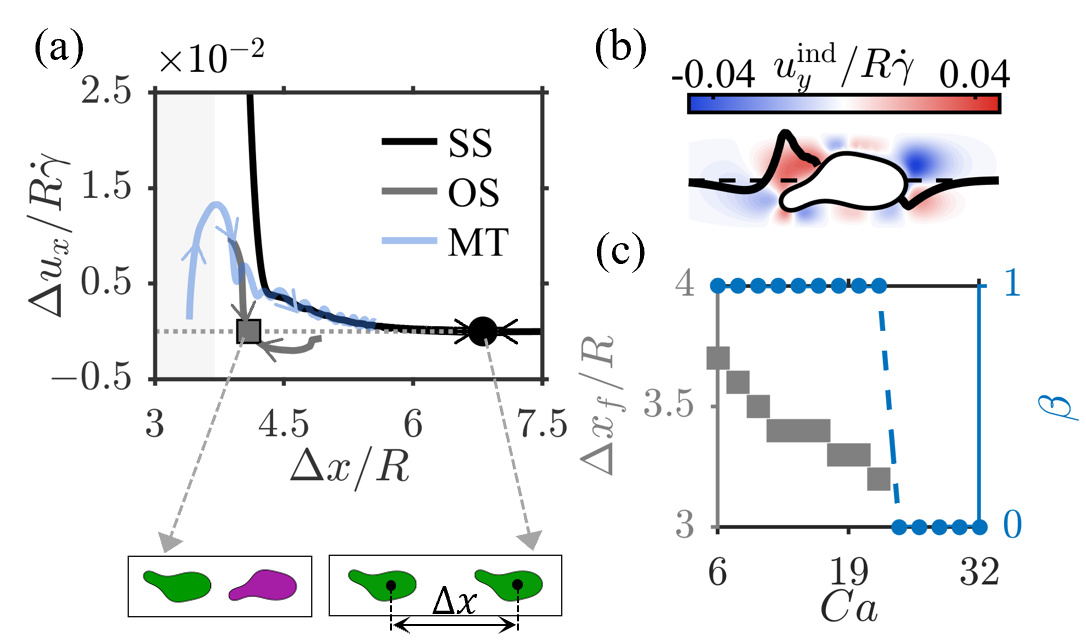}
\caption{\label{fig:fig2}
Directionally biased hydrodynamic interactions.
(a) Relative velocity between the two particles as a function of streamwise separation for same-sign (\(SS\)) and opposite-sign (\(OS\)) configurations at \(Ca=10\). The gray band marks the mode-transition (\(MT\)) window: when the pair enters this range, the \(OS\) state becomes unstable and switches to the \(SS\) branch.
(b) Transverse velocity field \(u^{\mathrm{ind}}_{y}\) generated by an isolated steady slipper at \(Ca=10\). The black curve shows the cross-sectionally averaged profile of \(u^{\mathrm{ind}}_{y}\).
(c) Critical forward-switching spacing \(\Delta x_f\) and directional-bias parameter \(\beta\) as functions of \(Ca\).
}
\end{figure}

\textit{Directionally biased switching of neighboring slippers}---To identify the microscopic origin of the front dynamics, we examine the interaction of two particles in Poiseuille flow using a sufficiently long channel to minimize periodic effects. The pair relaxes to a steady configuration with a characteristic spacing [Fig.~\ref{fig:fig2}(a)], and two branches are observed depending on the inclination signs. For same-sign pairs (\(SS\)), the equilibrium spacing is relatively large and the configuration is robust. For opposite-sign pairs (\(OS\)), the spacing is smaller and the configuration is more susceptible to finite compression.

Starting from an \(OS\) pair, we transiently force the particles closer together. The downstream slipper deforms and flips, whereas the upstream slipper retains its sign. The system thus leaves the \(OS\) branch and relaxes toward the \(SS\) state, after which the particles separate to the larger \(SS\) spacing [marked as \(MT\) in Fig.~\ref{fig:fig2}(a); see also Movie S2]. The switching is therefore one-sided: only the downstream particle changes branch. In this operational sense, the switching response is effectively directional.

This asymmetry originates from the fore-aft asymmetry of the slipper and the disturbance flow it generates. For an isolated downward slipper (\(S=-1\)), the disturbance field ahead of the particle carries a transverse advection toward \(-y\) [Fig.~\ref{fig:fig2}(b)] (the sign reverses for \(S=+1\)). When two particles approach, this front-side disturbance acts directly on the tail of the downstream slipper, which is the most compliant part of the shape. By contrast, the influence transmitted upstream comes mainly from the wake of the downstream slipper and acts first on the head of the upstream particle, which is less deformable. Under strong confinement, the rapid spatial decay of the disturbance field further weakens the perturbation reaching the upstream tail.

To quantify this directional bias, we measure two critical spacings for an \(OS\) pair forced closer together: \(\Delta x_f\), at which the downstream particle changes sign, and \(\Delta x_b\), the corresponding threshold for the upstream particle. We define \(\beta=\frac{1/\Delta x_b-1/\Delta x_f}{1/\Delta x_b+1/\Delta x_f}\), 
where a missing switching event is represented by \(\Delta x=0\), so that \(1/\Delta x\to\infty\). In this convention, \(\beta=1\) corresponds to a fully directional switching response, while \(\beta=0\) indicates that no bias is manifested. Over the range \(6\lesssim Ca\lesssim 32\), where an isolated particle exhibits stable slipper dynamics, we find \(\beta=1\) for \(Ca\lesssim 22\): the downstream particle switches at a finite \(\Delta x_f\), whereas the upstream particle does not switch even under the strongest compression. For larger \(Ca\), neither particle switches, and \(\beta=0\). This one-sided \(OS\to SS\) transition provides the elementary event that, when repeated along a train, gives rise to a streamwise switching front.

\textit{Propagating switching front}---In a train, consider a configuration \(\swarrow \nwarrow \nwarrow \nwarrow \cdots\), where the first slipper has a downward tail and all others have upward tails. When the leading \(OS\) pair (particles 1--2) undergoes an \(OS\to SS\) transition, the downstream particle flips and the pair relaxes to the larger \(SS\) spacing [Fig.~\ref{fig:fig2}(a)]. This rearrangement advects particle 2 toward particle 3, creating a new \(OS\) pair (particles 2--3), so that the train becomes \(\swarrow \swarrow \nwarrow \nwarrow \cdots\). If the resulting approach is sufficiently strong, the same transition repeats for the next pair, yielding \(\swarrow \swarrow \swarrow \nwarrow \cdots\), then \(\swarrow \swarrow \swarrow \swarrow \nwarrow \cdots\), and so on. In this way, successive local switching events relay the inclination sign along the train and generate a propagating switching front. The propagation stops when the flipped slipper meets a neighbor with the same sign, when the front reaches the end of the train, or when the transmitted perturbation is too weak to trigger switching of the next pair.

\begin{figure}
\centering
\includegraphics{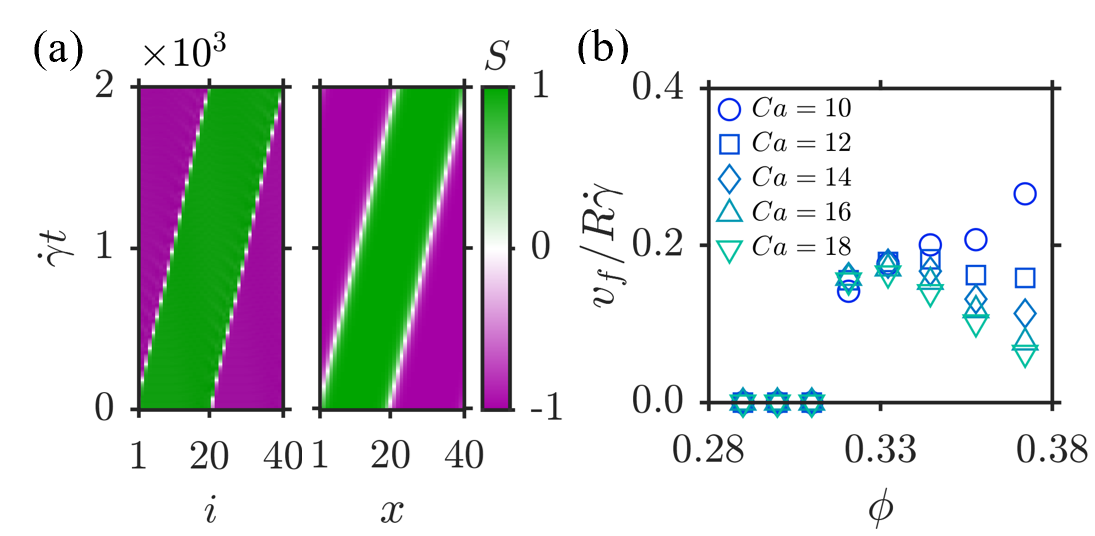}
\caption{\label{fig:fig3}
Propagating switching front.
(a) Space-time plot of the inclination order parameter \(S\) for a periodic train of \(N=40\) particles initialized with \(S=+1\) in the upstream half and \(S=-1\) in the downstream half, showing a persistent traveling switching front. Left: hydrodynamic simulations at \(Ca=20\) and \(\phi\simeq 32\%\). Right: minimal model [Eq.~(\ref{eq:RADE_S})] using effective parameters \(c\) and \(D\) for the same \((Ca,\phi)\).
(b) Front speed \(v_f\) as a function of \(\phi\) for different \(Ca\).
}
\end{figure}

We characterize this front dynamics in hydrodynamic simulations using a uniformly spaced periodic train (\(N=40\)), initialized with \(S=+1\) in the upstream half and \(S=-1\) in the downstream half [Fig.~\ref{fig:fig3}(a)]. Traveling fronts are observed within an intermediate concentration window, \(31\%\lesssim\phi\lesssim38\%\), and over a corresponding range of \(Ca\) [Fig.~\ref{fig:fig3}(b)]. Below this range, the front either fails to propagate or arrests, indicating that the effective coupling is insufficient to sustain successive switching events. Above it, the train itself becomes unstable, so that a sustained front can no longer be identified.

Within the propagation window, the front speed \(v_f\) is set by the slower of two processes: switching/shape relaxation and post-switch separation of the pair. Once an \(OS\to SS\) event is triggered, the downstream particle flips while the two particles move apart. At larger \(Ca\), switching and shape recovery become slower relative to the imposed flow, whereas the separation time scale depends mainly on \(\phi\). Accordingly, at small \(\phi\), \(v_f\) is separation-limited and only weakly dependent on \(Ca\) (\(\phi\lesssim 34\%\)) [Fig.~\ref{fig:fig3}(b)], while at larger \(\phi\) the dynamics becomes deformation-limited and \(v_f\) decreases with \(Ca\) (\(\phi\gtrsim 34\%\)) [Fig.~\ref{fig:fig3}(b)].

The onset of front propagation is also consistent with the two-particle threshold measured in Fig.~\ref{fig:fig2}(c). In the present geometry, a uniform spacing equal to the measured \(\Delta x_f\) corresponds to an area fraction of about \(33\%\) to \(38\%\) depending on \(Ca\) [Fig.~\ref{fig:fig2}(c)], whereas front transmission in the many-particle simulations is observed at about \(31\%\) [Fig.~\ref{fig:fig3}(b)]. This slight offset is physically expected: the two-particle estimate assumes a uniform spacing, whereas in a many-particle train switching is triggered by the local concentration. Local concentration fluctuations therefore allow neighboring pairs to enter the switching regime even when the mean area fraction lies slightly below the uniform two-particle threshold.

To rationalize the hydrodynamic front dynamics, we introduce a minimal model that retains only the two ingredients suggested by the simulations: local bistability of the inclination state and directional coupling along the train. The inclination order parameter \(s_i(t)\) of the \(i\)th particle, indexed from upstream to downstream, obeys
\begin{equation}
\dot{s}_i =
\mu s_i-s_i^{3}
+
C_1(s_{i-1}-s_i)
+
C_2(s_{i+1}-s_i),
\label{eq:discrete_model}
\end{equation}
where \(\mu s_i-s_i^3\) represents the local bistability and \(C_1\gg C_2\) encodes the stronger upstream-to-downstream influence inferred from the two-particle dynamics. Upon coarse graining, Eq.~(\ref{eq:discrete_model}) becomes
\begin{equation}
\partial_t s + c\partial_x s =
\mu s-s^3
+
D\partial_{xx}s,
\label{eq:RADE_S}
\end{equation}
with \(c=\Delta x(C_1-C_2)\) and \(D=\Delta x^2(C_1+C_2)/2\). Here \(c>0\) describes directional transmission of the inclination state, and \(D\) is an effective diffusion coefficient.

This reduced model is not intended as a parameter-free derivation of the full many-body hydrodynamics. Rather, it separates the front dynamics from the specific details of the hydrodynamic interactions and makes explicit the underlying mechanism: local bistability combined with directional coupling. It thus isolates the minimal ingredients required for front propagation, direction selection, and arrest. Equation~(\ref{eq:RADE_S}) reproduces the front dynamics when \(c\) and \(D\) are chosen as effective parameters for the corresponding \((Ca,\phi)\). Although explicit forms \(c(Ca,\phi)\) and \(D(Ca,\phi)\) cannot be derived from the full hydrodynamics, their variation is constrained by the simulated front propagation [Fig.~\ref{fig:fig3}(b)]. For example, at \(Ca=20\) and \(\phi\simeq 32\%\), the reduced model yields a front with the same propagation direction and speed as in the simulations [Fig.~\ref{fig:fig3}(a) and Movie S3].

\begin{figure}
\centering
\includegraphics{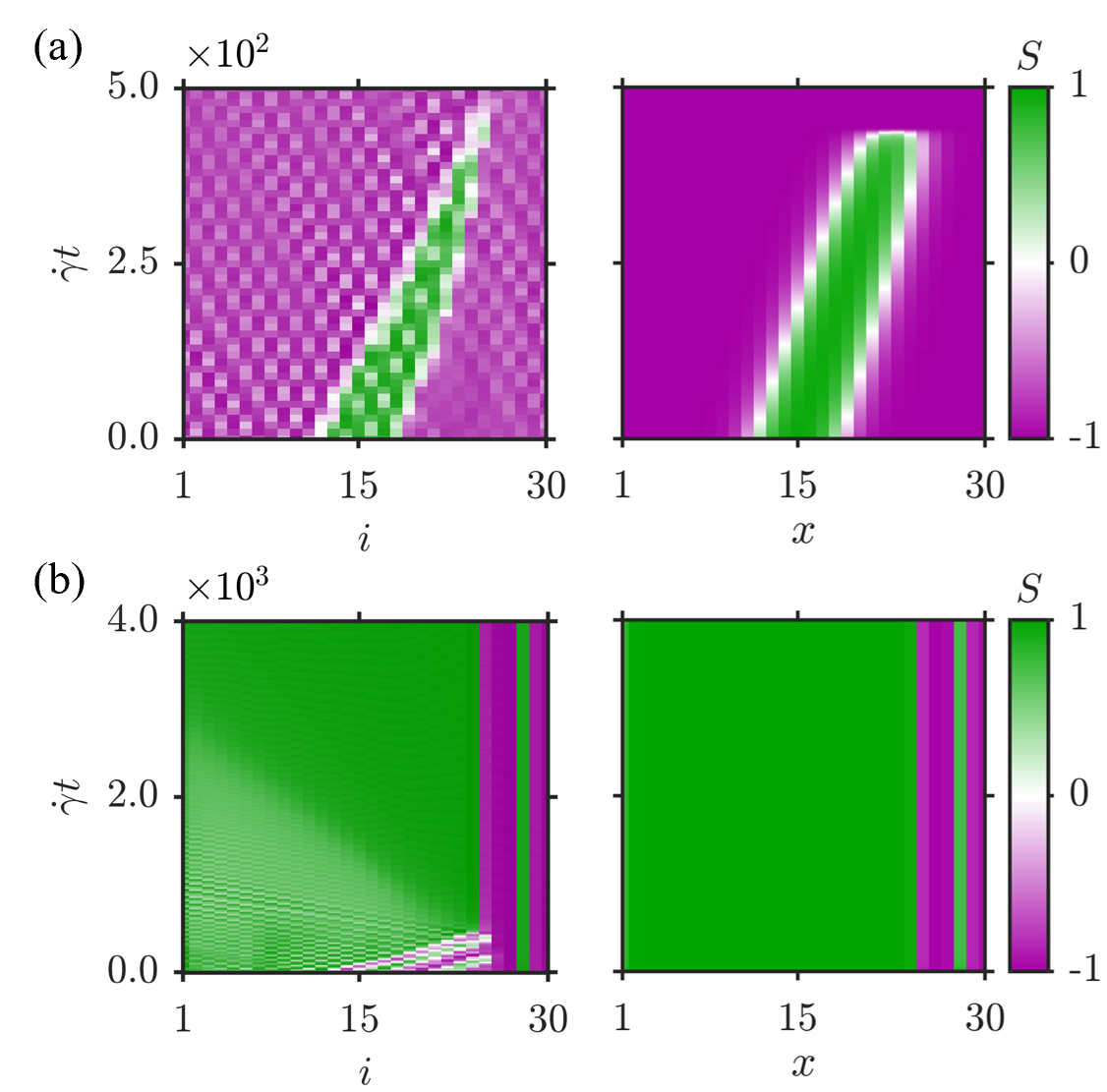}
\caption{\label{fig:fig4}
Coarsening and arrest of switching fronts.
(a) Two switching fronts with different propagation speeds, arising from spatial heterogeneity, drift toward each other, collide, and annihilate. Left: hydrodynamic simulations (same parameters as in Fig.~\ref{fig:fig1}). Right: minimal model.
(b) In a long channel, an effectively open train develops multiple polarized domains separated by \(OS\) pairs that act as domain boundaries. Left: hydrodynamic simulations (\(L=200R\), \(N=30\), \(\phi\simeq 20\%\), \(Ca=10\)). Right: minimal model with corresponding effective parameters.}
\end{figure}

\textit{Coarsening and arrest of switching fronts}---In the controlled two-domain initialization, a propagating front can persist for long times because of the imposed symmetry of the initial condition. By contrast, for random initial conditions we consistently observe complete polarization of the train, with all slippers adopting a common inclination sign. This difference arises because local heterogeneity in spacing and deformation produces fronts with unequal propagation speeds. As a result, neighboring fronts drift, collide, and annihilate [Fig.~\ref{fig:fig4}(a) and Movie S4]. In a periodic train, the number of fronts therefore decreases monotonically, and the dynamics coarsens toward a uniformly polarized state (Movie S1).

A different outcome is obtained in an open train, where each \(OS\to SS\) event increases the local interparticle spacing and thereby weakens the perturbation transmitted to the next pair. Once this perturbation falls below the switching threshold, the front arrests and stable \(OS\) pairs can persist. We observe the same qualitative scenario in long-channel simulations initialized from a locally condensed cluster. Because the channel is sufficiently long, periodic images remain remote and the evolving train is effectively open. At early times, the local concentration is high enough to sustain successive \(OS\to SS\) events [Fig.~\ref{fig:fig4}(b) left and Movie S5] and produce polarized segments. As the train relaxes and the mean spacing increases, the effective coupling weakens, further switching becomes unlikely, and the train settles into a steady multi-domain configuration with stable interfaces [Fig.~\ref{fig:fig4}(b) left and Movie S5]. The minimal model reproduces the same qualitative behavior when the effective coupling is reduced to values corresponding to sufficiently low concentration [Fig.~\ref{fig:fig4}(b) right].

\textit{Conclusions}---Our results show that collective polarity selection in confined trains of deformable particles can arise through propagating switching fronts. The key ingredient is a directionally biased transmission of inclination state, generated by the fore-aft asymmetry of the slipper under confinement. At the level of neighboring pairs, this asymmetry makes an \(OS\to SS\) transition more effective in the upstream-to-downstream direction than in the reverse direction. At the many-body level, it organizes successive switching events into a streamwise front, which coarsens into complete polarization in periodic trains but arrests into persistent domains in open trains. 

These findings provide a simple physical picture for how shape asymmetry and confinement produce directional collective dynamics in suspensions of deformable particles. More broadly, our results show that propagating fronts and collective polarity selection do not require activity or artificially imposed asymmetric couplings, but can arise in passive many-body systems when local bistability is combined with directionally biased state transmission. They may also provide insight into red-blood-cell train organization in microcirculation and suggest possible routes for the microfluidic control of deformable-particle assemblies.

\begin{acknowledgments}
We thank G. Li for useful discussions. Z.S. acknowledge National Natural Science Foundation of China (Grant No. 12572304) and the Natural Science Foundation of Beijing, China (Grant No. 1252020) for funding; Z.Z. acknowledge the Natural Science Foundation of Beijing, China (Grant No. 2262016) and 
National Natural Science Foundation of China (Grant No. 12404234) for funding.
\end{acknowledgments}


\bibliography{apssamp}

\end{document}